\documentclass[12pt]{iopart}
\usepackage{graphicx}
\usepackage{wrapfig}
\begin{document}

\title[{\footnotesize Polarization Reversal of Electron Cyclotron Wave  Due to Radial Boundary Condition}]{Polarization Reversal of Electron Cyclotron Wave Due to Radial Boundary Condition}

\author{K. Takahashi
\footnote[3]{kazunori@ecei.tohoku.ac.jp}
, T. Kaneko, and R. Hatakeyama}

\address{Department of Electronic Engineering, Tohoku University, Sendai 980-8579, Japan}


\begin{abstract}
\baselineskip 13pt
 Propagation and absorption of electromagnetic waves with electron cyclotron resonance (ECR) frequency are experimentally and theoretically investigated for the case of inhomogeneously magnetized plasma column with peripheral vacuum layer, when a left-hand polarized wave (LHPW) is selectively launched.
 The polarization reversal from the LHPW to the right-hand polarized wave is found to occur near the ECR point.
 As a result, it is clarified that the LHPW, which has been considered not to be absorbed at the ECR point, is absorbed near the ECR point.
 The phenomena can be explained by taking into account the effects of the radial boundary conditions.
 In addition, it is found that the polarization reversal point can be adjusted by the external parameters, for example, plasma radius.
\end{abstract}

\baselineskip 14pt
\section{Introduction}
\hspace{4ex}
 Electron cyclotron wave is important plasma waves in the fields of basic plasma physics, nuclear fusion, and plasma processing.
 In recent years, suppression of neoclassical tearing modes and current drive in tokamak are influentially attempted by using electron cyclotron resonance (ECR) in the field of nuclear fusion \cite{Prater1}, where the localized electron heating plays important roles in.
 The localized electron cyclotron resonance heating is hoped to be the most effective method for the formation of confining-potential structure in tandem-mirror devices \cite{Kaneko1,Tamano1}.
 In the field of plasma processing, on the other hand, productions of ECR plasma with large diameter, high-density, and uniformity were reported \cite{Ueda1}.
 ECR has attractive characteristics and many applications as mentioned above.
 Although only a right-hand polarized wave (RHPW) has been considered to be absorbed near the ECR point, the recent experimental results demonstrated that a left-hand polarized wave (LHPW) was also efficiently and locally absorbed [4-6].
 We have suggested the polarization reversal from the LHPW to the RHPW as an origin of the unexpected absorption of the LHPW\cite{Kaneko2}, and recently succeeded in observation of the polarization reversal\cite{Takahashi1}.
 The detailed mechanisms of the polarization reversal is worth clarifying in the terms of basic plasma physics and localized electron heating.\par
 Based on these background, the purposes of the present work are to clarify the polarization-reversal mechanisms and characteristics.
\par

\section{Experimental Setup}
\hspace{4ex}
 Experiments are performed in the Q$_{\rm{T}}$-Upgrade Machine of Tohoku University as shown in Fig.~1(a).
 A coaxial bounded plasma with peripheral vacuum layer is produced by a direct current discharge in low pressure argon gas (90 mPa).
 The plasma column is terminated on the glass endplate located on the opposite side of the plasma source.
 The formation of clear boundary between the plasma and the vacuum layer is realized by using limiter, which is located on the front of tungsten mesh anode and also can control the plasma radius $r_p$ in the range of $2.5 - 4\ \rm{cm}$.
 The typical electron density and temperature at the radial center of the plasma column are $n_e\simeq 9\times 10^{10}\ \rm{cm^{-3}}$ and $T_e = 3\ \rm{eV}$.
 Static magnetic-field $B$ configuration is inhomogeneous as presented in Fig.~1(b).
 The LHPW (6 GHz, 150 mW) is selectively launched in the high magnetic-field region by a helical antenna ($z = 0\ \rm{cm}$), where the ECR point ($B=2.14\ \rm{kG}$) is $z=78\ \rm{cm}$.
 The wave patterns are obtained with an interference method through movable dipole antennas, which can receive the each components of wave electric field, i.e., $E_x$, $E_y$, and $E_z$, respectively.
 Moreover, spatial profiles of the wave power $P_x$, $P_y$, and $P_z$ are measured with a power meter.
\par
\vspace{5mm}
\setcounter{figure}{1}
\begin{figure}[htbp]
\begin{center}
  	\includegraphics[width=.65\linewidth]{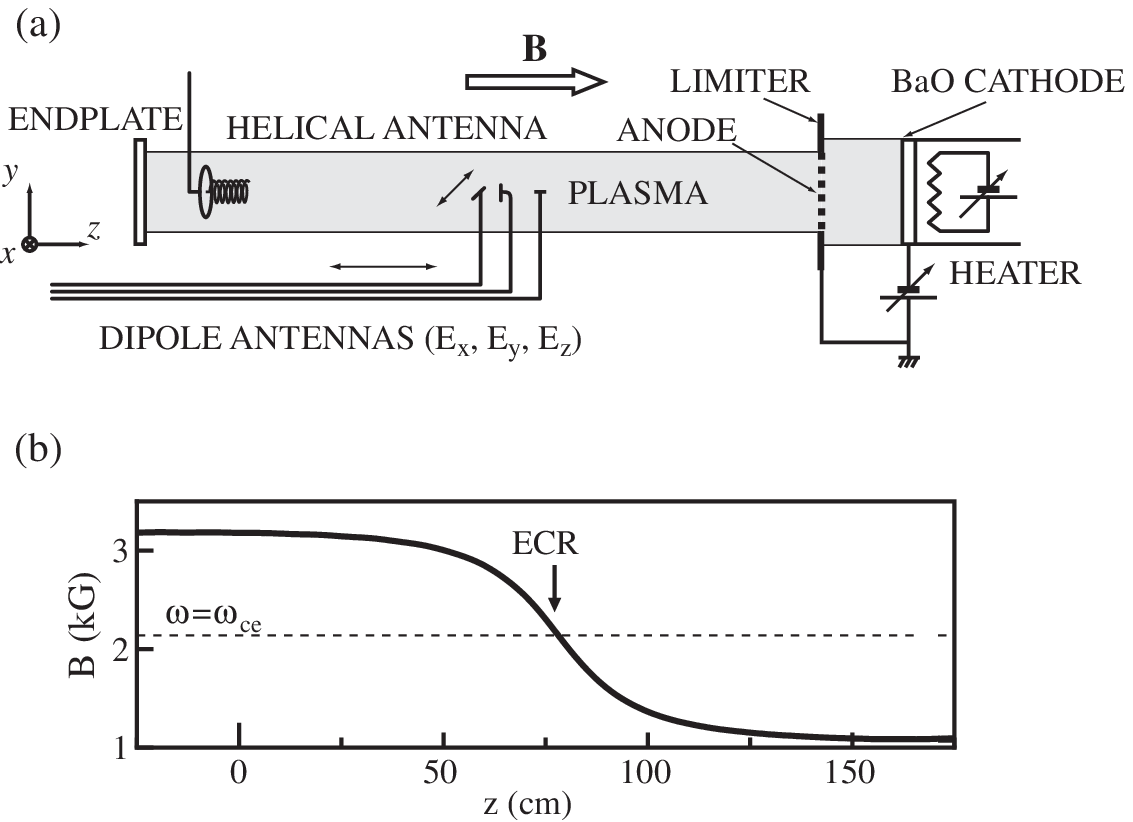}\\
\baselineskip 13pt
{\small	{\bf Fig.~1} : (a)Schematic of experimental apparatus. (b)Static magnetic-field configuration.}
\end{center}
\end{figure}

\section{Experimental and Theoretical Results}
\hspace{4ex}
\begin{figure}[htbp]
 \begin{center}
  \includegraphics[width=.45\linewidth]{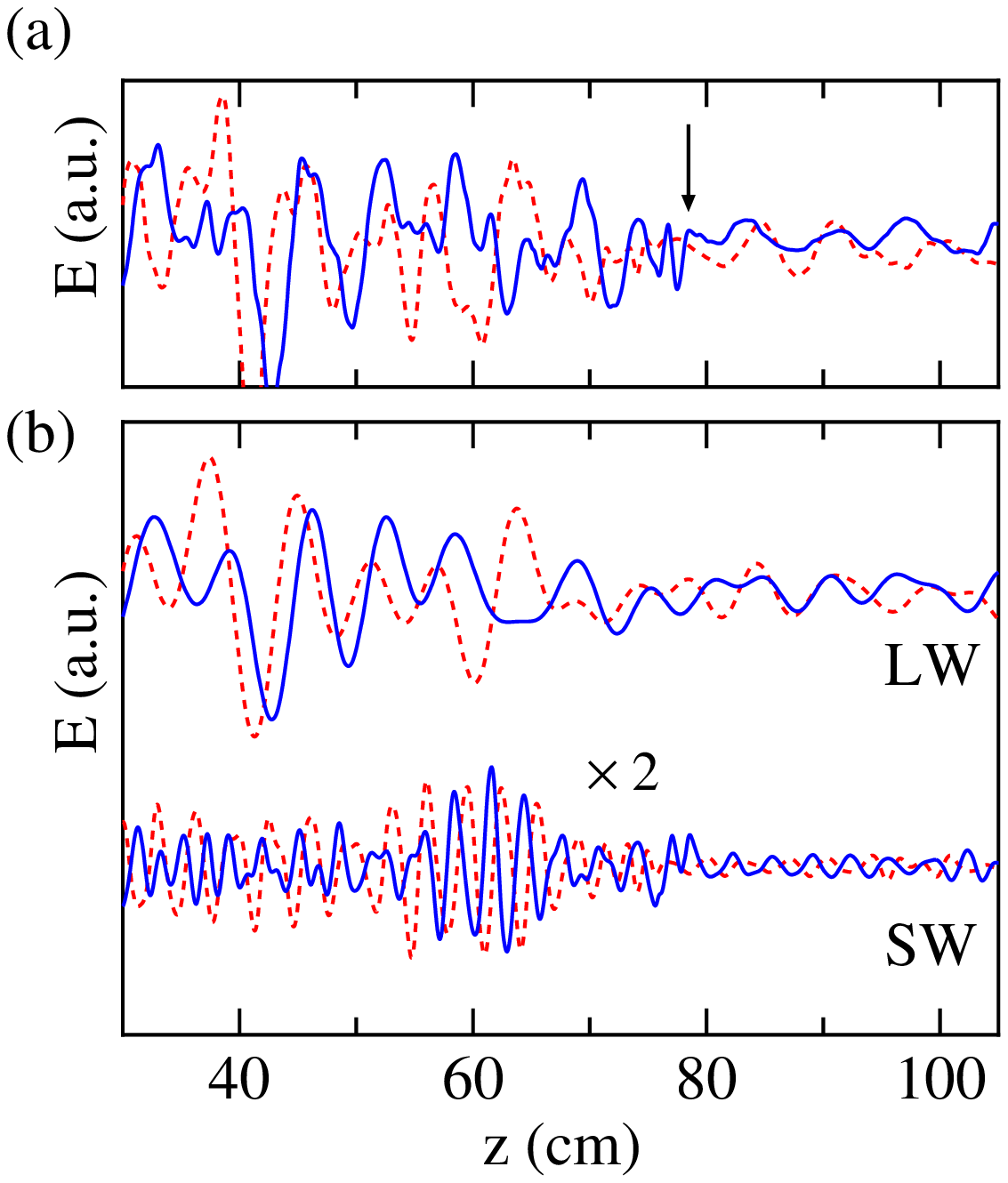}\\
 \label{WavePatterns}
\baselineskip 13pt
{\small	{\bf Fig.~2} : (a)Interferometric wave patterns of $E_x$ (dashed line) and $E_y$ (solid line). (b)Long \mbox{(LW) and short (SW) wavelength components decomposed from the wave patterns in Fig}.~2(a).}
 \end{center}
\end{figure}

 Figure~2(a) shows the interferometric wave patterns of $E_x$ (dashed line) and $E_y$ (solid line) at the radial center of the plasma column, where the ECR point is indicated by arrow.
 The wave patterns also demonstrate the unexpected absorption of the LHPW near the ECR point as reported in previous paper \cite{Kaneko2}.
 The long (LW) and short (SW) wavelength components, which are decomposed from the observed wave patterns in Fig.~2(a) using Fourier analysis, are presented in Fig.~2(b).
 The wave patterns of $E_x$ in the LW and the SW are shifted to the left and to the right of the wave patterns $E_y$, respectively.
 From these phase differences between $E_x$ and $E_y$, the LW and SW are identified as the LHPW and the RHPW, respectively.
 In addition, it is noticed that the LHPW damps and the RHPW grows around $z=60\ \rm{cm}$; namely, the polarization reversal from the LHPW to the RHPW is experimentally evidenced near the ECR point.
 As a result of the polarization reversal, the launched LHPW is found to be absorbed near the ECR point.\par
 We apply the dispersion relation of electromagnetic wave in bounded plasmas, which includes the effects of the radial boundary condition, in order to explain the polarization reversal and to evaluate the observed wave patterns.
 The dispersion equation is derived by using the Maxwell equations as \cite{Swanson}
\begin{center}
\begin{eqnarray}
\vspace{2mm}
(\gamma^2 + \kappa_2^2 + \gamma k_{\perp}^2)\kappa_3 + k_{\perp}^2[\kappa_1(\gamma+k_{\perp}^2)-\kappa_2^2] = 0,
\vspace{2mm}
\label{DispersionEq}
\end{eqnarray}
\end{center}
where $\gamma \equiv k_{\parallel}^2 - \kappa_1$, $k_{\parallel}$ and $k_{\perp}$ are the wave number in parallel and perpendicular to a static magnetic field ${\bf B}$.
 $\kappa_1$, $\kappa_2$, and $\kappa_3$ are the dielectric tensor elements and defined as
\begin{center}
\begin{eqnarray}
\hspace{-5ex}\frac{c^2}{\omega^2} \kappa_1 &=& 1-\sum_j\frac{\omega_{pj}^2}{\omega^2-\omega_{cj}^2},\nonumber\\
\hspace{-5ex}i\frac{c^2}{\omega^2} \kappa_2 &=& \sum_j\frac{\epsilon_j\omega_{cj}\omega_{pj}^2}{\omega(\omega^2-\omega_{cj}^2)},\nonumber\\
\hspace{-5ex}\frac{c^2}{\omega^2} \kappa_3 &=& 1-\sum_j\frac{\omega_{pj}^2}{\omega^2}
\end{eqnarray}
\end{center}
where $\omega_{pj}$, $\omega_{cj}$, and $\epsilon_j$ are the plasma frequency, the cyclotron frequency, and the sign of the charge for species $j$.
 In addition, the dispersion relation described by Eq.~(\ref{DispersionEq}) derive a polarization index $S$ as
\begin{center}
\begin{eqnarray}
\hspace{-5ex}S \equiv \frac{|E_r + iE_{\theta}|}{|E_r - iE_{\theta}|}.\hspace{9ex}
\label{Polarization-index}
\end{eqnarray}
\end{center}
Here, $0<S<1$ and $1<S<\infty$ represent right-handed and left-handed polarizations, and $S=0$, $S=1$, and $S=\infty$ correspond to circularly right-handed, linear, and circularly left-handed polarizations.
 The wave frequency $\omega/2\pi$ is constant and the static magnetic-field strength $B$ is varied when we consider the spatial wave propagation in $z$-direction.
 Therefore, not wave frequency $\omega/2\pi$ but the magnetic-field strength $B$ has to be regarded as a variable in calculating the dispersion relation given by Eq.~(\ref{DispersionEq}) for inhomogeneous magnetic-field configuration.
\par
 The calculated dispersion curve ($k_{\parallel}$ vs. $B$) is presented in Fig.~3(a) as a solid line together with the experimentally obtained dispersion relation of the LHPW (closed square) and the RHPW (open square), where the dashed lines in Figs.~3(a) and 3(b) denote the magnetic-field strength at the ECR point.
 The experimentally obtained dispersion relation of the LHPW and the RHPW have a good agreement with the calculated one.
 Thus, it is found that the wave propagation can be well described by including the effect of the radial boundary condition.
 The calculated polarization index $S$ as a function of magnetic-field strength $B$ is plotted in Fig.~3(b), which corresponds to the dispersion curve in Fig.~3(a).
 When the wave propagates from high magnetic-field region to the ECR point, it is found that the value of $S$ turns from $S>1$ into $S<1$, namely, the wave polarization is varied from left-handed to right-handed.
 Therefore, the observed polarization reversal can be also explained by including the effect of the radial boundary.\par
 We carry out a detailed investigation about the effects of the radial boundary on the wave propagation and polarization reversal, where the plasma radius is changed in the range of $r_p=2.5-4\ \rm{cm}$.
 The absorption of the LHPW and the polarization reversal are observed for any plasma radius.
\begin{figure}[htbp]
\begin{center}
  	\includegraphics[width=.9\linewidth]{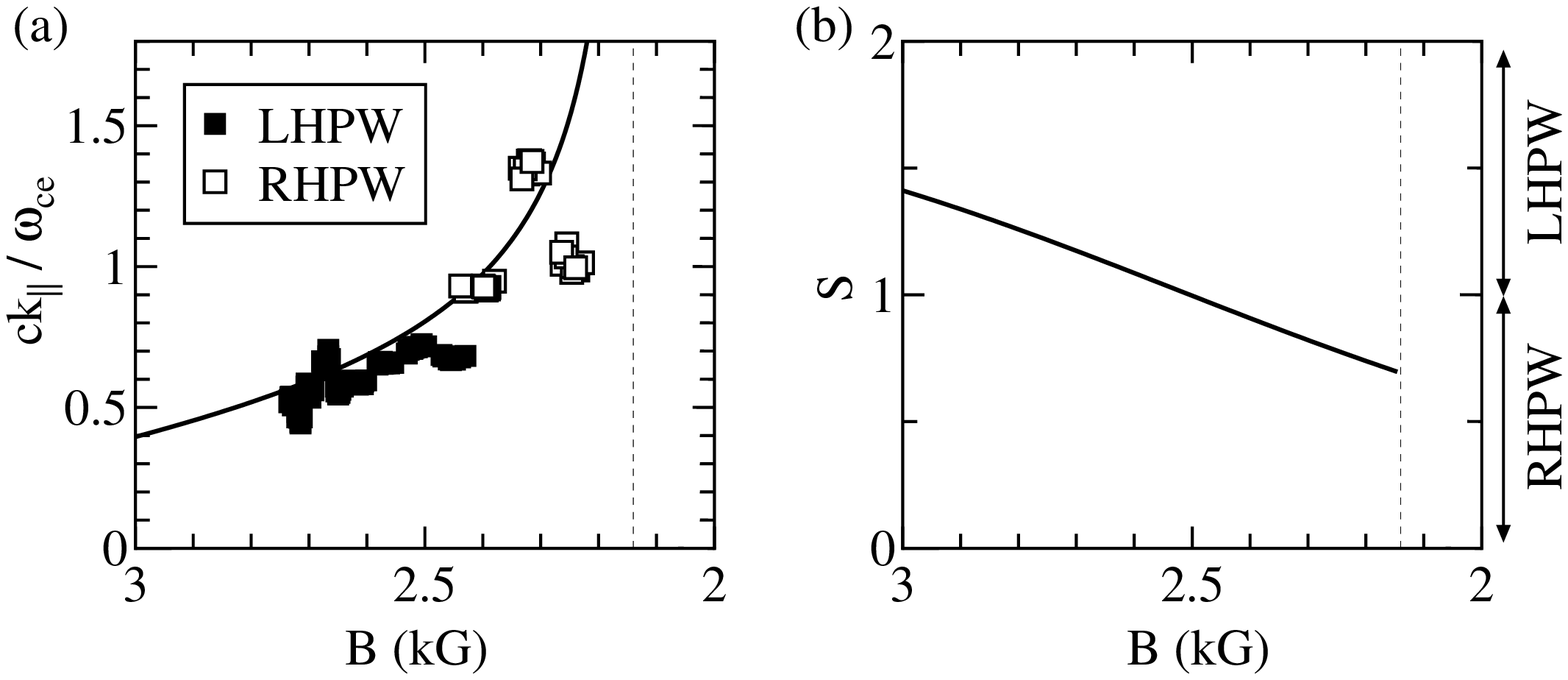}\\
	\label{Dispersion}
\end{center}
\baselineskip 13pt
{\small	{\bf Fig.~3} : (a)Calculated dispersion relation (solid line) together with the experimental dispersion relation of the LHPW (closed square) and the RHPW (open square). (b)Calculated polarization index $S$ corresponding to the dispersion curve in Fig.~3(a).}
\vspace{-5mm}
\end{figure}
 It is found that the polarization reversal point, where the RHPW components has a maximum amplitude, is shifted with changing the plasma radius.
 Figure~4(a) shows the experimentally observed magnetic-field strength $B_{pr}$ at the polarization reversal point as a function of the plasma radius $r_p$.
 The polarization reversal point is found to move to high magnetic-field region with an increase in $r_p$.
 This phenomenon is found to be also explained by the dispersion relation in bounded plasmas.
 Figure~4(b) shows the calculated polarization indexes ($B$ vs. $S$) for $r_p=2.5$, $3$, and $4\ \rm{cm}$, where the arrows indicate the magnetic-field strength at the polarization reversal point ($S=1$).
 The magnetic-field at the polarization reversal point is found to become higher with an increase in $r_p$, which is consistent with the experimentally observed phenomenon as shown in Fig.~4.
 These results implies that the polarization reversal point could be controlled by adjusting the plasma radius, which yields the control of the wave-absorption region near the ECR point.

\begin{figure}[htbp]
\begin{center}
  	\includegraphics[width=.85\linewidth]{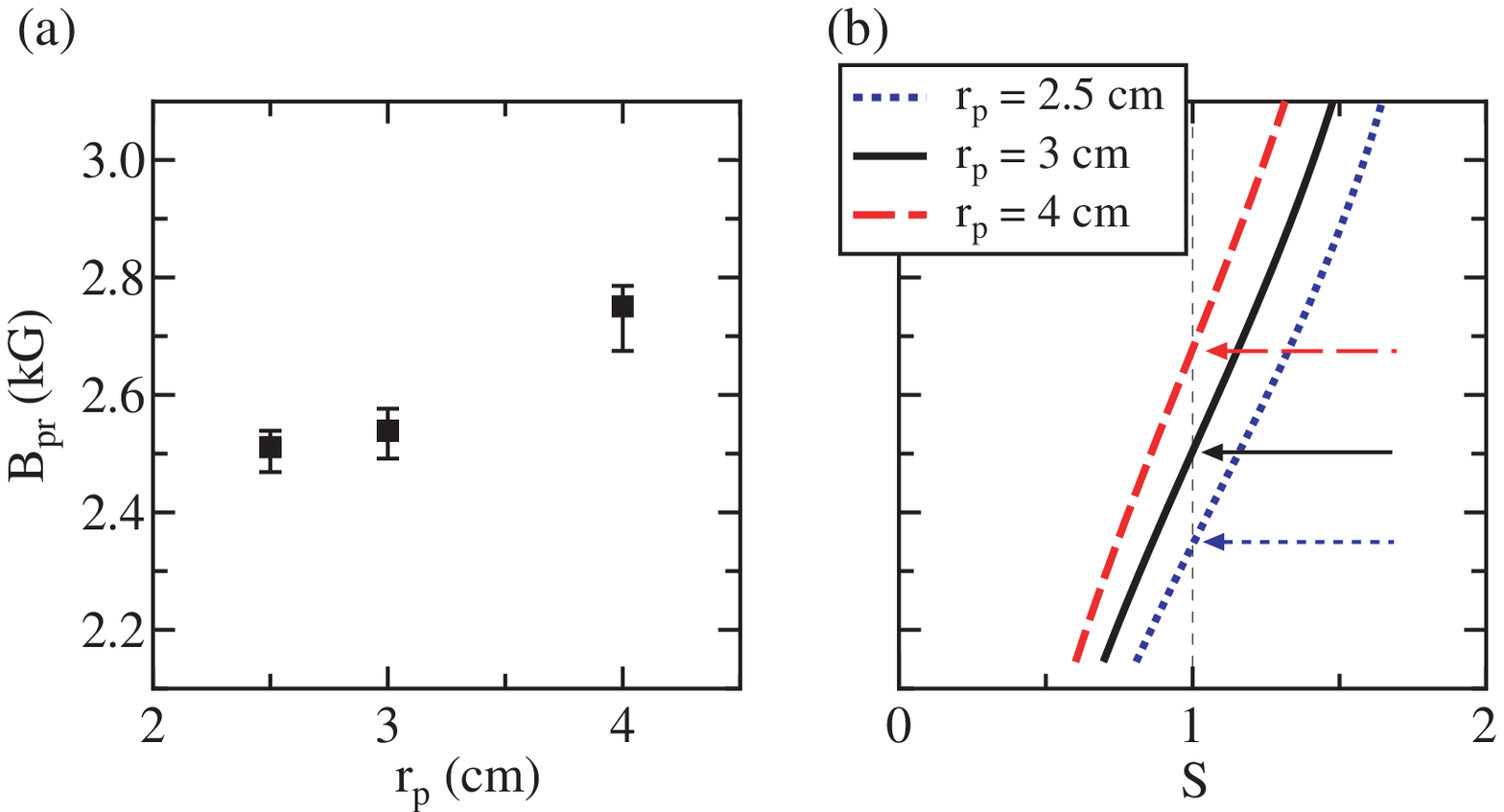}\\
	\label{PRpoint}
\end{center}
\baselineskip 13pt
{\small	{\bf Fig.~4} : (a)Experimentally observed magnetic-field strength at the polarization reversal point as a function of plasma radius $r_p$. (b)Calculated polarization index $S$ ($B$ vs. $S$) for $r_p = 2.5$ (dotted line), $3$ (solid line), and $4\ \rm{cm}$ (dashed line). The arrows denote the calculated magnetic-field strength at the polarization reversal point for each $r_p$.}
\end{figure}

\section{Conclusion}
 The propagation and absorption of the electromagnetic wave near the ECR point are investigated for the case of inhomogeneously magnetized plasma, when the left-hand polarized wave (LHPW) is selectively launched by using the helical antenna.
 Our experimental results demonstrate the polarization reversal from the LHPW to the right-hand polarized wave (RHPW) and the resultant absorption of the LHPW.
 The polarization reversal can be well explained by the dispersion relation including the effect of the radial boundary.
 In addition, it is found that the polarization reversal point is shifted by the plasma radius.
 This phenomenon can be interpreted in the term of the dispersion relation in bounded plasmas.
 Here, the possibility of the polarization reversal point is implied, which could play important role in the control of the wave absorption region near the ECR point.
 
\section*{Acknowledgement}
The authors are indebted to H. Ishida for his technical assistance. The work was supported by a Grant-in-Aid for Scientific Research from the Ministry of Education, Culture, Sports, Science and Technology, Japan.

\end{document}